\documentclass[11pt]{article}

\usepackage[utf8]{inputenc}
\usepackage[T1]{fontenc}
\usepackage{amsmath,amssymb,amsfonts,bm}
\usepackage{geometry}
\usepackage{booktabs}
\usepackage{graphicx}
\usepackage{authblk}
\usepackage[numbers,square,sort&compress]{natbib}
\usepackage[colorlinks=true,linkcolor=blue,citecolor=blue,urlcolor=blue]{hyperref}

\title{The Bioelectrical Information Theory: Investigating the theoretical compression limit of bioelectrical signals under artificial intelligence}

\author[1]{Jiawen Zou}
\author[1]{Bo Yan\textsuperscript{\dag}}
\affil[1]{College of Computer Science and Artificial Intelligence, Shanghai Key Laboratory of Intelligent Information Processing, Fudan University, Shanghai 200433, P.R. China\\
\textsuperscript{\dag}Corresponding author: \texttt{byan@fudan.edu.cn}}

\date{}

\begin{document}
\maketitle

\begin{abstract}
    Bioelectrical signals are increasingly acquired at scales that challenge the bandwidth of brain-computer interfaces. However, their compression is still often framed as a problem of waveform preservation, limited by the entropy of the raw signal. Here we propose an information-theoretic framework in which the effective information of bioelectrical data is determined not only by signal fidelity, but also by physiological structure, model capacity and downstream task requirements. We formulate bioelectrical compression as a three-level hierarchy. At the signal level, noise is reduced to the information they carry about latent physiological sources. At the physiological level, parametric encoders map purified signals into compact, structured and quantized representations. At the semantic level, task-irrelevant information is discarded, while deep learning models exploit causal dependencies to replace marginal entropy with conditional entropy. This perspective reframes the compression limit of bioelectrical signals as a model- and task-conditioned quantity rather than a fixed property of the waveform. As increasingly expressive models become integrated with neural and physiological interfaces, bioelectrical compression may shift from transmitting signals to transmitting only the residual information required for task-level interpretation.
\end{abstract}

\section{Introduction}

Bioelectrical signals provide a real-time interface between physiological dynamics and computational systems, enabling neural communication prostheses, wearable health monitoring, and closed-loop bioelectronic control \cite{willett2021high,metzger2023high,shin2022wearable}. However, the scale of modern acquisition is beginning to exceed the bandwidth, storage, and energy budgets of practical sensing platforms. High-density neural probes and cortical microelectrode arrays can now record from thousands of sites over long time horizons \cite{steinmetz2021neuropixels,hettick2025minimally}, while embedded neural interfaces increasingly treat compression as part of the acquisition front end to reduce data movement \cite{jang20231024}. Compression is therefore not merely an engineering convenience for bioelectrical sensing, but a prerequisite for deploying physiological intelligence under constrained power, latency, and communication budgets \cite{du2024fast,jang20231024}.

Existing approaches have compressed bioelectrical signals for waveform reconstruction \cite{ille2024ongoing,jang20231024,du2024fast}, compact feature extraction \cite{LaBraM,VQ-MTM,sani2021modeling}, and downstream task such as semantic prediction \cite{alemi2016deep,dubois2020learning,abbaspourazad2023large,li2025electrocardiogram}. This progression, from signal reconstruction to model-based representation learning and semantic compression, calls for a corresponding re-examination of the effective information of bioelectrical data. Here we formulate bioelectrical signal compression as a three-level hierarchy. At the signal level, raw signals are treated as noisy mixtures of latent physiological sources, and their effective information content is determined by the mutual information between the observation and the underlying source activity. At the physiological level, parametric encoders compress purified signals into structured, low-dimensional representations that preserve salient physiological dynamics. At the semantic level, task-relevant information is separated from semantic nuisance components, and models such as large language models or world models further exploit causal and contextual dependencies in structured physiological representations to replace marginal entropy with conditioned entropy. This hierarchy suggests that the compression limit of bioelectrical signals is not fixed by waveform fidelity alone, but by the interaction between physiological structure, model priors, and downstream task requirements, thus allowing extreme entropy reduction while preserving task-relevant information.

\section{Main}

\begin{figure}[t]
    \centering
    \includegraphics[width=\linewidth]{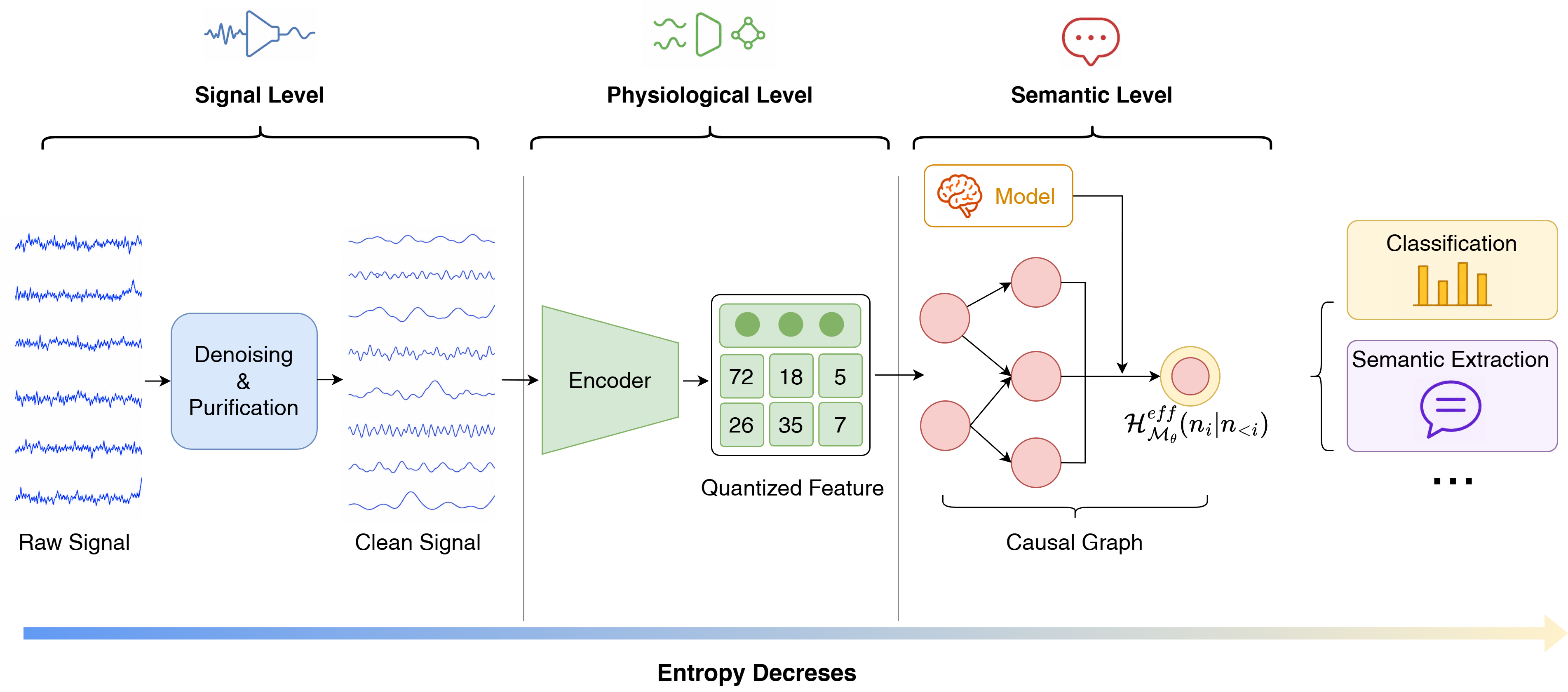}
    \caption{
    \textbf{Hierarchical information reduction for bioelectrical signal compression.}
    Raw bioelectrical recordings first undergo signal-level purification, where noise and measurement noise are removed to recover cleaner physiological signals. The purified signals are then mapped by an encoder into compact quantized features, yielding a structured physiological representation with reduced dimensionality and finite resolution. At the semantic level, the structured representation can be regarded as a causal graph, in which upstream contextual variables provide the conditions for predicting downstream units. The model, such as a large language model or world model, removes components that are irrelevant to downstream objectives and exploits causal context to encode the conditional entropy rather than marginal entropy. The resulting semantic code preserves information required by downstream objectives, such as  classification and semantic extraction, etc.
    }
    \label{fig:main_framework}
\end{figure}

As illustrated in Fig.~\ref{fig:main_framework}, we organize bioelectrical signal compression as a hierarchical process of information reduction rather than a single-stage waveform coding problem. The method proceeds through three progressively levels. First, signal-level purification removes measurement noise, retaining the information that the observed recording carries about latent physiological sources. Second, physiological layer maps the purified signal into a compact, structured, and quantized representation that preserves salient physiological dynamics while reducing dimensionality and resolution. Third, semantic layer extracts task-relevant information from the structured representation and models the causal dependencies to further reduce the entropy.

\paragraph{Signal-level purification.}

The first layer of compression removes noise from the raw waveform. We model the observed bioelectrical signal as a noisy superposition of the latent physiological source activity. Let $Z_{raw}\in\mathbb{R}^{m}$ denote the observed multi-channel signal and $Z_{sig}$ denote the latent clean source. The signal-level observation can be modeled as:
\begin{equation}
    Z_{raw} = A Z_{sig} + N,
    \label{eq:signal_model}
\end{equation}
where $A\in\mathbb{R}^{m\times m}$ is a mixing operator such as volume conduction, channel mixing, convolutional spreading, or other linearized forward effects, and $N\in\mathbb{R}^{m}$ denotes noise, such as sensor noise and background activity. Under the Gaussian assumptions $Z_{sig}\sim\mathcal{N}(0,\Sigma_{Z_{sig}})$, $N\sim\mathcal{N}(0,\Sigma_N)$, and $Z_{sig}\perp N$, the effective source information contained in the noisy observation is given by the Gaussian vector-channel mutual information \cite{cover1999elements,telatar1999capacity}:
\begin{equation}
    I(Z_{raw};Z_{sig})
    =
    h(Z_{raw})-h(Z_{raw}\mid Z_{sig})
    =
    \frac{1}{2}
    \log_2
    \det
    \left(
        I+
        \Sigma_N^{-\frac{1}{2}}
        A\Sigma_{Z_{sig}}A^\top
        \Sigma_N^{-\frac{1}{2}}
    \right).
    \label{eq:signal_effective_information}
\end{equation}
This quantity measures the part of the raw signal that is informative about latent physiological activity, rather than the total entropy of the measurement itself.

\paragraph{Physiological-level abstraction.}

After signal-level purification, the second layer compresses the clean source representation $Z_{sig}$ into a compact physiological representation $Z_{phys}$. The objective is no longer to preserve the complete waveform, but to retain the salient structure needed to describe the underlying physiological dynamics. This stage can be implemented by a parametric encoder, such as a vector-quantized autoencoder \cite{van2017neural,VQ-MTM}, which maps the source signal into a lower-dimensional and discrete representation. Under a high-resolution quantization approximation \cite{gray2002quantization}, the corresponding entropy can be written as
\begin{equation}
    H(Q_{\delta_{phys}}(Z_{phys}))
    \approx
    h(Z_{phys})
    +
    d_{phys}\log_2\frac{1}{\delta_{phys}}
    <
    h(Z_{sig})
    +
    d_{sig}\log_2\frac{1}{\delta_{sig}}
    \approx
    H(Q_{\delta_{sig}}(Z_{sig})),
    \label{eq:physio_entropy_approx}
\end{equation}
where $Q_{\delta_{sig}}(\cdot)$ and $Q_{\delta_{phys}}(\cdot)$ denote quantization with resolutions $\delta_{sig}$ and $\delta_{phys}$, respectively; $d_{sig}$ and $d_{phys}$ denote the effective dimensions of the signal-level and physiological-level representations; and $h(\cdot)$ denotes differential entropy. Thus physiological-level compression can reduce entropy through two mechanisms: lowering the intrinsic representation dimension $d$, such as feature engineering \cite{ramoser2000optimal}, and coarsening the value resolution $\delta$, such as quantized representation learning \cite{VQ-MTM}. The achievable reduction is therefore governed by the ability of the encoder-decoder pair to preserve salient physiological structure while discarding signal-level redundancy.

\paragraph{Semantic-level conditioning.}

At the physiological level, bioelectrical signals are compressed into structured, low-dimensional, and quantized representations, such as sequences of discrete tokens \cite{VQ-MTM}. The final layer maps the structured representation $Z_{phys}$ into a task-relevant semantic code $Z_{sem}$. Unlike physiological-level compression, which preserves signal morphology and temporal dynamics, semantic-level compression is concerned with the information in $Z_{phys}$ that is useful for a downstream task. Therefore, the first role of semantic-level compression is to discard semantic nuisance components that are not informative for the task. For a task $\mathcal{T}$ and the corresponding semantic information $Y_{\mathcal{T}}$, the task-relevant effective information in $Z_{phys}$ can be written as
\begin{equation}
    H^{eff}(Z_{phys})
    \triangleq
    I(Z_{phys};Y_{\mathcal{T}})
    =
    H(Z_{phys}) - H(Z_{phys}\mid Y_{\mathcal{T}})
    <
    H(Z_{phys}).
    \label{eq:task_effective_information}
\end{equation}
This expression emphasizes that semantic compression does not need to preserve all variability in $Z_{phys}$; it only needs to preserve the part that is informative about $Y_{\mathcal{T}}$.

The second role of semantic-level compression is to exploit the causal or contextual dependencies embedded in $Z_{phys}$. Structured physiological representation's components are not independent symbols but relies on the corresponding context. To encode the structured representation $Z_{phys}=Z({n_1,n_2,\ldots,n_k})$, conventional encoding schemes such as Huffman coding mainly exploit marginal symbol statistics. This marginal coding baseline can be written in a compact form as
\begin{equation}
    H^{eff}(Z_{phys})
    \sim
    \sum_i H^{eff}_{\mathrm{context}}(n_i)
    \label{eq:huffman_effective_information}
\end{equation}
where context contains shared prior knowledge such as the Huffman tree. Although classical methods can use limited context, they are usually not expressive enough to capture long-range causal dependencies among physiological components. Modern deep learning models (denoted by $\mathcal{M_\theta}$), especially large language models or world models, provide a more powerful expression capability to model such dependencies. Conditioning on the model and context, the marginal entropy can be reduced to conditional entropy:
\begin{equation}
    H(Z_{sem})
    =
    H^{eff}_{\mathcal{M_\theta}}(Z_{phys})
    =
    \sum_i
    H^{eff}_{\mathcal{M_\theta}}
    \left(
        n_i
        \mid
        n_{<i}
    \right)
    <
    \sum_i H^{eff}_{\mathrm{context}}(n_i)
    \label{eq:model_conditioned_effective_information}
\end{equation}
Thus, semantic-level compression reduces the effective entropy in two ways: it removes information irrelevant to $Y_{\mathcal{T}}$, and uses $\mathcal{M_\theta}$ to model causal dependencies to reduce the marginal entropy to conditional entropy.

\paragraph{Overall entropy decomposition.}

The three levels above provide a staged view of how bioelectrical information can be progressively reduced. Signal-level purification keeps the part of the observation that is informative about the latent source; physiological-level abstraction maps the purified source into a compact structured representation; and semantic-level conditioning retains only task-relevant information and exploiting model-based causal predictability to model conditional entropy. As a conceptual approximation, the total entropy reduction can be decomposed as
\begin{equation}
\begin{aligned}
-\Delta H
& \approx
\underbrace{
\left[
H(Z_{raw})
-
\frac{1}{2}
\log_2
\det
\left(
I+
\Sigma_N^{-\frac{1}{2}}
A\Sigma_{Z_{sig}}A^\top
\Sigma_N^{-\frac{1}{2}}
\right)
\right]
}_{\text{signal-level purification}}
\\
&\quad+
\underbrace{
\left[
(h(Z_{sig}) + d_{sig}\log_2\frac{1}{\delta_{sig}})
-
(h(Z_{phys}) + d_{phys}\log_2\frac{1}{\delta_{phys}})
\right]
}_{\text{physiological-level abstraction}}
\\
&\quad+
\underbrace{
\left[
H(Z_{phys}\mid Y_{\mathcal{T}})
+
\sum_i H^{eff}_{\mathrm{context}}(n_i) - 
\sum_i H^{eff}_{\mathcal{M_\theta}}
\left(
    n_i
    \mid
    n_{<i}
\right)
\right]
}_{\text{semantic-level relevance and conditioning}} .
\end{aligned}
\label{eq:overall_entropy_decomposition}
\end{equation}

Equation~\eqref{eq:overall_entropy_decomposition} should be interpreted as a conceptual decomposition rather than a strict additive law. The three layers may interact, their reductions need not be statistically independent, and the optimal representation at one level can depend on the task and model priors used at another. Therefore, the bracketed terms should not be understood as an exact telescoping decomposition. Nevertheless, Eq.~\eqref{eq:overall_entropy_decomposition} captures the central intuition of the proposed framework: bioelectrical compression can proceed from waveform purification, to physiological abstraction, and finally to semantic task resolution. In this view, modern large language models and world models lower the effective entropy not by reconstructing the signal more accurately, but by making more of the structured representation predictable from causal-semantic context.

\section{Conclusion}

In this work, we have presented an information-theoretic view of bioelectrical signal compression as a hierarchical process of information reduction. Rather than treating bioelectrical recordings as raw waveforms to be faithfully reconstructed, this formulation separates their information content into three progressively more abstract levels: signal-level purification, physiological-level abstraction, and semantic-level conditioning. At the signal level, compression removes noise and retains the information that the noisy observation carries about latent physiological activity. At the physiological level, parametric encoders further reduce the representation by extracting compact, structured, and quantized features. At the semantic level, the objective shifts from reconstructing the physiological representation itself to preserving the information that is relevant to a downstream task.

This perspective suggests that the compression limit of bioelectrical signals is not determined solely by the marginal entropy of the recorded waveform. Instead, the relevant information becomes the residual uncertainty that remains after conditioning on physiological structure, causal dependencies, model priors, and task requirements. In particular, large language models and world models provide a mechanism for exploiting causal and contextual regularities in structured physiological representations, allowing predictable components to be inferred rather than transmitted. This changes the role of compression from signal preservation to semantic coding, where only the information needed to update a shared structure toward the correct task-level interpretation is communicated. As large language models and world models become increasingly integrated with bioelectrical sensing platforms, model-aware compression may enable substantially lower entropy while preserving the information required for high-level communications.

\end{document}